\documentclass[10pt,twocolumn,letterpaper]{article}

\usepackage[pagenumbers]{cvpr} % To force page numbers, e.g. for an arXiv version

\usepackage{times}
\usepackage{epsfig}
\usepackage{graphicx}
\usepackage{amsmath}
\usepackage{amssymb}

\usepackage{bbm}
\usepackage{bm}
\usepackage{xspace}
\usepackage[utf8]{inputenc} % allow utf-8 input
\usepackage{url}            % simple URL typesetting
\usepackage{booktabs}       % professional-quality tables
\usepackage{amsfonts}       % blackboard math symbols
\usepackage{nicefrac}       % compact symbols for 1/2, etc.
\usepackage{microtype}      % microtypography
\usepackage{color}
\usepackage{algorithm}
\usepackage{algorithmic}
\usepackage{array}
\usepackage{multirow}
\usepackage{enumitem}
\usepackage{makecell}
\usepackage{gensymb}         % for degree
\usepackage{mathtools}
\usepackage{dblfloatfix}
\usepackage{pifont}
\usepackage[permil]{overpic}

\usepackage{titletoc,tocloft}

\newcommand{\cmark}{\ding{51}}

\makeatletter
\DeclareRobustCommand\onedot{\futurelet\@let@token\@onedot}
\def\@onedot{\ifx\@let@token.\else.\null\fi\xspace}
\def\eg{\emph{e.g}\onedot} 
\def\ie{\emph{i.e}\onedot}

\makeatother

\usepackage{xcolor}
\usepackage{ifthen}
\usepackage{textcomp}
\definecolor{red}{rgb}{1,0,0}
\definecolor{slateblue}{rgb}{0.7,0.35,0.9}
\definecolor{green}{rgb}{0,1,0}
\definecolor{mahogany}{rgb}{0.75, 0.25, 0.0}
\definecolor{purple}{rgb}{0.6, 0, 0.6}
\definecolor{darkpurple}{rgb}{0.3, 0, 0.3}
\definecolor{darkgreen}{rgb}{0, 0.4, 0}
\definecolor{frenchblue}{rgb}{0.0, 0.45, 0.73}
\definecolor{blue}{rgb}{0,0,1}
\definecolor{goldenrod}{rgb}{0.65, 0.45, 0.03}
\definecolor{gray}{rgb}{0.6,0.6,0.6}
\definecolor{gold}{rgb}{1.0, 0.874, 0}
\definecolor{silver}{rgb}{0.67,0.67,0.67}
\definecolor{brown}{rgb}{0.8, 0.678, 0.4}

\newcommand{\gray}[1]{\textcolor{gray}{#1}}
\newcommand{\gold}[1]{\colorbox{gold}{\textbf{#1}}}
\newcommand{\silver}[1]{\colorbox{silver}{\textbf{#1}}}
\newcommand{\brown}[1]{\colorbox{brown}{\textbf{#1}}}

\newboolean{revising}
\setboolean{revising}{true}
\ifthenelse{\boolean{revising}}
{
    \newcommand{\ignore}[1]{}

} {
    \newcommand{\ignore}[1]{}

}

\makeatletter
\renewcommand{\paragraph}{%
  \@startsection{paragraph}{4}%
  {\z@}{0.5\baselineskip \@plus 0ex \@minus 0ex}{-1em}%
  {\normalfont\normalsize\bfseries}%
}
\makeatother

\makeatletter
\newcommand\footnoteref[1]{\protected@xdef\@thefnmark{\ref{#1}}\@footnotemark}
\makeatother

\usepackage{listings}
\definecolor{codegreen}{rgb}{0,0.6,0}
\definecolor{codegray}{rgb}{0.5,0.5,0.5}
\definecolor{codepurple}{rgb}{0.58,0,0.82}
\definecolor{backcolour}{rgb}{0.95,0.95,0.92}
\lstdefinestyle{mystyle}{
    backgroundcolor=\color{backcolour},   
    commentstyle=\color{codegreen},
    keywordstyle=\color{magenta},
    numberstyle=\tiny\color{codegray},
    stringstyle=\color{codepurple},
    basicstyle=\ttfamily\footnotesize,
    breakatwhitespace=false,         
    breaklines=true,                 
    captionpos=b,                    
    keepspaces=true,                 
    numbers=left,                    
    numbersep=5pt,                  
    showspaces=false,                
    showstringspaces=false,
    showtabs=false,                  
    tabsize=2
}
\usepackage[pagebackref,breaklinks,colorlinks]{hyperref}

\usepackage[capitalize]{cleveref}
\crefname{section}{Sec.}{Secs.}
\Crefname{section}{Section}{Sections}
\Crefname{table}{Table}{Tables}
\crefname{table}{Tab.}{Tabs.}

\def\cvprPaperID{90} % *** Enter the CVPR Paper ID here
\def\confName{}
\def\confYear{}

\begin{document}

%%%%%%%%% TITLE
\title{Improved Direct Voxel Grid Optimization for Radiance Fields Reconstruction}

\author{
Cheng Sun \qquad Min Sun \qquad Hwann-Tzong Chen \\
National Tsing Hua University \\
{\tt\small chengsun@gapp.nthu.edu.tw} \qquad {\tt\small sunmin@ee.nthu.edu.tw} \qquad {\tt\small htchen@cs.nthu.edu.tw}
}

\maketitle
%\thispagestyle{empty}

%%%%%%%%% ABSTRACT
\begin{abstract}
In this technical report, we improve the DVGO~\cite{SunSC22} framework (called DVGOv2), which is based on Pytorch and uses the simplest dense grid representation.
First, we re-implement part of the Pytorch operations with cuda, achieving $2$--$3\times$ speedup.
The cuda extension is automatically compiled just in time.
Second, we extend DVGO to support Forward-facing and Unbounded Inward-facing capturing.
Third, we improve the space time complexity of the distortion loss proposed by mip-NeRF 360~\cite{BarronMVSH22} from $\mathcal{O}(N^2)$ to  $\mathcal{O}(N)$.
The distortion loss improves our quality and training speed.
Our efficient implementation could allow more future works to benefit from the loss.

% We hope DVGOv2 could be a good option as the starting point for future research, especially when there are other challenges to deal with.

Project page: \url{ https://sunset1995.github.io/dvgo/}.

Code: \url{ https://github.com/sunset1995/DirectVoxGO}.
\end{abstract}

\section{Introduction}
% NeRF and its computation issue
Neural radiance fields~\cite{MildenhallSTBRN20} (NeRF) have provided an appealing approach to novel view synthesis for the high quality and flexibility to reconstruct the volume densities and view-dependent colors from multi-view calibrated images.
However, NeRF runs very slow due to the processing time of multilayer perceptron (MLP) networks.
Consider an MLP consisting of 8 layers with 256 hidden channels: To query a single point would require more than $520k \approx 256^2 \cdot 8$ FLOPs.
In each training iteration, there are typically 8{,}192 rays, each with 256 sampled points, which results in more than 1T FLOPs in total.

% Occupancy grid
Using an occupancy mask is one of the easiest ways to speed up.
As the training progresses for a scene, we can gradually update an occupancy mask to deactivate some of the space with low density.
Thanks to the occupancy mask, we can then skip most of the point queries in each iteration after training for a while.
VaxNeRF~\cite{KondoITMOG21} reports that a vanilla NeRF with the occupancy mask can achieve $2$--$8\times$ speedups for bounded inward-facing scenes~\cite{KondoITMOG21}.
Most of the recent works~\cite{SunSC22,YuFTCR22,KondoITMOG21,ChenXGYS22,mueller2022instant} on training time speedup use the occupancy mask trick.

\begin{table}[t]
    \centering
    \begin{tabular}{@{}l|lll@{}}
    \hline
    Method & Data structure & Density & Color \\
    \hline\hline
    DVGO~\cite{SunSC22} & dense grid & explicit & hybrid \\
    Plenoxels~\cite{YuFTCR22} & octree & explicit & explicit \\
    Instant-NGP~\cite{mueller2022instant} & hash table & hybrid & hybrid \\
    TensoRF~\cite{ChenXGYS22} & decomposed grid & explicit & hybrid \\
    \hline
    \end{tabular}
    \vspace{-.5em}
    \caption{{\bf Overview of explicit radiance field representations.} Various data structures have been realized to model the volume densities and view-dependent colors explicitly. The `hybrid' indicates that the explicit representation is followed by an implicit representation (the MLP network).}
    \label{tab:recent_grid}
    \vspace{-1.5em}
\end{table}

% Recent progress using explicit representation
Recently, many works (\cref{tab:recent_grid}) have emerged using explicit representations to reduce training time from hours to minutes per scene.
Querying an explicit representation requires only constant time computation, which is much more efficient than the few hundred thousand FLOPs per query.
Even hybrid representations may benefit from the reduced computation for the speedup in training since the MLP network in a hybrid representation is typically much shallower than that in a fully implicit representation.
% Many data structures have been realized for the explicit representation, from the simplest dense grid to the advanced octree, hash-table, and decomposed components.
DVGO~\cite{SunSC22} uses the simplest dense grid data structure in fully Pytorch implementation.
Plenoxels~\cite{YuFTCR22} model the coefficients of spherical harmonic for view-dependent colors and realize a fully explicit (without MLP) representation.
Plenoxels interpolation and rendering pipeline are fused in CUDA code.
Instant-NGP~\cite{mueller2022instant} uses hash-table and hybrid representations for both densities and colors.
Instant-NGP further improves the training time using C/C++ and fully-fused CUDA implementation.
TensoRF~\cite{ChenXGYS22} improves the memory footprint and scalability of the dense grid via tensor decomposition and directly modeling the low-rank components.

This technical report presents DVGOv2.
Compared to DVGO, DVGOv2 achieves another $2$--$3\times$ speedup and extends to forward-facing and unbounded inward-facing capturing.
We also present an efficient realization of the distortion loss (reduced from $\mathcal{O}(N^2)$ to $\mathcal{O}(N)$), which improves our quality and training time.
DVGOv2 uses the simplest data structure and most of our intermediate steps are in Python interface.
Meanwhile, DVGOv2 still demonstrates the good quality and convergence speed.
% We believe DVGOv2 is easy to extend and incorporate with other methods, 
% , and iterate the newly 
% , which could be easy to extend and incorporate with other methods.
% , which allows developers to iterate the methods quickly with .

%%%%%%%%%%%%%%%%%%%%%%%%%

\section{Efficient regularization}

Unlike the implicit MLP representation, the explicit representation is found to be more prone to producing artifacts of holes or floaters~\cite{YuFTCR22}.
Thus, regularization losses are especially important for achieving reasonable results on the unbounded real-world captured scenes.

\paragraph{Efficient total variation (TV) loss.}
TV loss~\cite{RudinO94} is commonly adapted to prevent unnecessary sharpness in explicit modeling~\cite{YuFTCR22,ChenXGYS22}.
For each grid point, we compute the Huber loss to its six nearest-neighbor grid points; we find that the Huber loss is better than L1 and L2 loss in our case.

However, computing the TV loss is time-consuming for a large dense grid and requires many Pytorch API calls to implement, so we fuse them into a single CUDA kernel.
Besides, we skip the forward pass and directly add the gradient into the Pytorch tensor, so the users have to call our API between the normal backward pass and the optimization step (see \cref{lst:tv_loss}).
Despite the CUDA extension, it still takes a lot of time, so we only compute the TV loss densely for all grid points in the first 10k iterations.
After the 10k checkpoint, we only compute the TV loss for grid points involved in the current iteration (\ie, with non-zero gradients).

\begin{lstlisting}[language=Python,label={lst:tv_loss},caption=Call our efficient TV loss after the backward pass.]
optimizer.zero_grad()
# compute total_loss
total_loss.backward()
dvgo_model.total_variation_add_grad(
    tv_weight, dense_mode=(curr_step<10000))
optimizer.step()
\end{lstlisting}

\paragraph{Efficient $\mathcal{O}(N)$ distortion loss.}
The distortion loss is proposed by mip-NeRF 360~\cite{BarronMVSH22}.
For a ray with $N$ sampled points, the loss is defined as
\begin{multline} \label{eq:dist_loss}
    \mathcal{L}_{\text{dist}}(s,w) = \sum_{i=0}^{N-1} \sum_{j=0}^{N-1} w_i w_j \left| \frac{s_i + s_{i+1}}{2} - \frac{s_j + s_{j+1}}{2} \right| \\
        + \frac{1}{3} \sum_{i=0}^{N-1} w_i^2 \left(s_{i+1} - s_{i}\right) ~,
\end{multline}
where $(s_{i+1}{-}s_{i})$ is the length and $(s_i{+}s_{i+1})/2$ is the midpoint of the $i$-th query interval.
The $s$ is non-linearly normalized (from the near-far clipping distance to $[0, 1]$) to prevent overweighting the far query.
The weight $w_i$ is for the $i$-th sample points.
Despite we are using the point-based instead of the interval-based query (which is an interesting problem), we find it still beneficial to adapt the distortion loss.

However, the straightforward implementation for the first term in \cref{eq:dist_loss} results in $\mathcal{O}(N^2)$ computation for a single ray.
This is not a problem for mip-NeRF 360 as there are only $32$ query intervals in the finest sampling.
For the point-based query, there are typically more than $256$ sampled points on each ray, which makes the computation non-trivial and consumes many GPU memory (more than 3G for a batch with $4{,}096$ rays).

Thus, we re-implement the first term to achieve $\mathcal{O}(N)$ computation.
Let the mid-point distance $m_i=(s_i{+}s_{i+1})/2$ and $m_i < m_{i+1}$.
We can eliminate the diagonal term ($i=k, j=k$) and rewrite it into:
\begin{equation}
\footnotesize
\begin{split}
    \mathcal{L}_{\text{dist}}^{\text{1st}} &=\sum_{i=0}^{N-1} \sum_{j=0}^{N-1} w_i w_j \left| m_i - m_j \right| \\
    &= 2 \sum_{i=1}^{N-1} \sum_{j=0}^{i-1} w_i w_j (m_i - m_j) \\
    &= \left(2 \sum_{i=1}^{N-1} w_i m_i\sum_{j=0}^{i-1} w_j\right) - \left(2 \sum_{i=1}^{N-1} w_i \sum_{j=0}^{i-1}w_j m_j\right) ~,
\end{split}
\end{equation}
where we can compute and store the prefix sum of $(w)$ and $(w \odot m)$ first so we can directly lookup the results of the inner summation when computing the outer summation.
The overall computation can thus be realized in $\mathcal{O}(N)$.

The derivative for $w_k$ is
\begin{equation}
\footnotesize
\begin{split}
    &\frac{\partial}{\partial w_k} \mathcal{L}_{\text{dist}}^{\text{1st}} \\
    &= 2 \frac{\partial}{\partial w_k} \sum_{j=0}^{k-1} w_k w_j (m_k - m_j) + 2 \frac{\partial}{\partial w_k} \sum_{i=k+1}^{N-1} w_i w_k (m_i - m_k) \\
    &= 2 \sum_{j=0}^{k-1} w_j (m_k - m_j) + 2 \sum_{i=k+1}^{N-1} w_i (m_i - m_k) \\
    &= 2 m_k \sum_{j=0}^{k-1} w_j - 2 \sum_{j=0}^{k-1} w_j m_j + 2 \sum_{i=k+1}^{N-1} w_i m_j - 2 m_k \sum_{i=k+1}^{N-1} w_i ~,
\end{split}
\end{equation}
where we can also compute and store the prefix and suffix sum of $(w)$ and $(w\odot m)$ so we can directly lookup the result when computing the derivative of every $w_k$.
The overall computation is also $\mathcal{O}(N)$.

We implement the efficient distortion loss as Pytorch CUDA extension and support uneven number of sampled points on each ray.
We provide a self-contained package at
\url{https://github.com/sunset1995/torch_efficient_distloss}.
We will see that the distortion loss improves our rendering quality and speed up our training, thanks to the compactness encouraged by the loss.
We believe the efficient distortion loss and our implementation can let more works benefit from mip-NeRF 360's regularization technique as most NeRF-based methods have hundreds of sampled points on a ray.

%%%%%%%%%%%%%%%%%%%%%%%%%

\section{CUDA speedup}
There are lots of sequential point-wise operations in DVGO, each of which has an overhead for launching the CUDA kernel.
So we re-implement these sequential point-wise operations into a single CUDA kernel to reduce launching overhead.
We refer interested reader to \url{https://pytorch.org/tutorials/advanced/cpp_extension.html}.
We use Pytorch's just-in-time compilation mechanism, which automatically compiles the newly implemented C/C++ and CUDA code by the first time it is required.

\paragraph{Re-implement Adam optimizer.}
There are about ten point-wise operations in an Adam optimization step.
We fuse them into a single kernel and skip updating the grid points with zero gradients.

\paragraph{Re-implement rendering utils.}
Originally, we sample an equal number of points on each ray for vectorized Pytorch implementation, where a large number of query points are outside the scene BBox.
We now infer the ray BBox intersection to sample query points parsimoniously for each ray (which is only applicable to bounded scene).
Besides, we fuse about ten point-wise operations for the forward and the backward pass of the density to alpha function.
In the volume rendering accumulation procedure, we halt tracking a ray once the accumulated transmittance is less than $10^{-3}$.

%%%%%%%%%%%%%%%%%%%%%%%%%

\section{Mix of factors affecting speed and quality}
Please note that the comparisons on the training speed and the result quality in this report are affected by many factors, not just the different scene representations as presented in \cref{tab:recent_grid}.

First, the computation devices are different.
As shown in \cref{tab:gpu_specs}, the computing power across different works is not aligned, where we use the lowest spec GPU to measure our training times.
Second, Instant-NGP's training pipeline is implemented in C++, while the other methods use Python/Pytorch.
Instant-NGP and Plenoxels implement most of their computations (\eg, grid interpolation, ray-casting, volume rendering) in CUDA; DVGOv2 customizes part of the computation as CUDA extension, and most of the intermediate steps are still in Python interface; DVGO and TensoRF only use the built-in Pytorch API.
Third, some implementation details such as regularization terms, policy of occupancy grid, and the other tricks can affect the quality and convergence speed as well.

\begin{table}[h!]
    \centering
    \begin{tabular}{@{}l|c@{\hskip 6pt}cl@{}}
    \hline
    GPU & FLOPs & Memory & Used by \\
    \hline\hline
    RTX 2080Ti & 13.45T & 11G & DVGO~\cite{SunSC22}, DVGOv2 \\
    Telsa V100 & 15.67T & 16G & TensoRF~\cite{ChenXGYS22} \\
    Titan RTX  & 16.31T & 24G & Plenoxels~\cite{YuFTCR22} \\
    RTX 3090   & 35.58T & 24G & Instant-NGP~\cite{mueller2022instant} \\
    \hline
    \end{tabular}
    \vspace{-0.5em}
    \caption{\small{\bf GPU specs.} FLOPs are theoretical for float32.}
    \label{tab:gpu_specs}
    \vspace{-1em}
\end{table}

%%%%%%%%%%%%%%%%%%%%%%%%%

\section{Experiments}

\subsection{Ablation study for the CUDA speedup}
We test the re-implementation with $160^3$ voxels on the {\it lego}, {\it mic}, and {\it ship} scenes.
The PSNRs of different version are roughly the same.
We present the results in \cref{tab:cuda_reimpl}, where the training time is measured on an RTX 2080Ti GPU and is $2$--$3\times$ faster than the original implementation.
We use the improved implementation in the rest of this technical report.

\begin{table}[h!]
    \centering
    \begin{tabular}{@{}cc@{\hskip 3pt}|@{\hskip 3pt}c@{\hskip 3pt}c@{\hskip 3pt}|@{\hskip 3pt}c@{\hskip 3pt}c@{\hskip 3pt}|@{\hskip 3pt}c@{\hskip 3pt}c@{}}
    \hline
    Adam & Rendering & \multicolumn{2}{c@{\hskip 3pt}|@{\hskip 3pt}}{{\it lego}} & \multicolumn{2}{c@{\hskip 3pt}|@{\hskip 3pt}}{{\it mic}} & \multicolumn{2}{c@{}}{{\it ship}} \\
    \hline\hline
    & & 11.5m & & 9.3m & & 14.6m & \\
    \hline
    \cmark & & 8.7m & 1.3x & 6.4m & 1.5x & 12.1m & 1.2x \\
    \cmark & \cmark & {\bf 4.8m} & {\bf 2.4x} & {\bf 3.4m} & {\bf 2.7x} & {\bf 7.1m} & {\bf 2.1x} \\
    \hline
    \end{tabular}
    \vspace{-.5em}
    \caption{{\bf Speedup by the improved implementation.}}
    \label{tab:cuda_reimpl}
    % \vspace{-1em}
\end{table}

\subsection{Bounded inward-facing scenes}
We evaluate DVGOv2 on two bounded inward-facining datasets---Synthetic-NeRF~\cite{MildenhallSTBRN20} and Tanks\&Temples~\cite{KnapitschPZK17} dataset (bounded ver.).
The results are summarized in \cref{tab:bd_inward}.
DVGOv2's training time is two more times faster than DVGO.
DVGOv2 also uses less training time than most of the recent methods despite using the lowest spec GPU.
The result qualities are also comparable to the recent methods.
The improvement by scaling to a higher grid resolution is limited on the Tanks\&Temples~\cite{KnapitschPZK17} dataset, perhaps because of the photometric variation between training views.

\begin{table}[!h!]
    \centering

    \begin{subtable}{\linewidth}
    \centering
    \begin{tabular}{@{}l@{\hskip 3pt}|@{\hskip 3pt}c@{\hskip 8pt}c@{\hskip 8pt}c@{\hskip 8pt}c@{}}
    \hline
    Method & Tr. time & PSNR{\scriptsize$\uparrow$} & SSIM{\scriptsize$\uparrow$} & LPIPS{\scriptsize(VGG)$\downarrow$} \\
    \hline\hline
    DVGO~\cite{SunSC22} & 14.2m & 31.95 & 0.957 & 0.053 \\
    Plenoxels~\cite{YuFTCR22} & \gray{11.1m} & 31.71 & \brown{0.958} & \brown{0.049} \\
    Instant-NGP~\cite{mueller2022instant} & \gray{5m} & \gold{33.18} & - & - \\
    TensoRF (S)~\cite{ChenXGYS22} & \gray{13.9m} & 32.39 & 0.957 & 0.057 \\
    TensoRF (L)~\cite{ChenXGYS22} & \gray{8.1m}  & 32.52 & 0.959 & 0.053 \\
    TensoRF (L)~\cite{ChenXGYS22} & \gray{17.6m} & \silver{33.14} & \gold{0.963} & \silver{0.047} \\
    DVGOv2 (S) & 4.9m & 31.91 & 0.956 & 0.054 \\
    DVGOv2 (L) & 6.8m & \brown{32.76} & \silver{0.962} & \gold{0.046} \\
    \hline
    \end{tabular}
    \caption{\small {\bf Results on Synthetic-NeRF~\cite{MildenhallSTBRN20} dataset.}
    The results are averaged over 8 scenes.}
    \label{tab:bd_nerf_summary}
    \end{subtable}

    \par\medskip

    \begin{subtable}{\linewidth}
    \centering
    \begin{tabular}{@{}l@{\hskip 3pt}|@{\hskip 3pt}c@{\hskip 8pt}c@{\hskip 8pt}c@{\hskip 8pt}c@{}}
    \hline
    Method & Tr. time & PSNR{\scriptsize$\uparrow$} & SSIM{\scriptsize$\uparrow$} & LPIPS{\scriptsize(VGG)$\downarrow$} \\
    \hline\hline
    DVGO~\cite{SunSC22} & 17.7m & \brown{28.41} & \brown{0.911} & \brown{0.155} \\
    TensoRF (S)~\cite{ChenXGYS22} & - & 28.06 & 0.909 & 0.155 \\
    TensoRF (L)~\cite{ChenXGYS22} & - & \silver{28.56} & \gold{0.920} & \gold{0.140} \\
    DVGOv2 (S) & 7.3m & 28.29 & 0.910 & 0.157 \\
    DVGOv2 (L) & 9.1m & \gold{28.69} & \silver{0.918} & \silver{0.143} \\
    \hline
    \end{tabular}
    \caption{\small {\bf Results on Tanks\&Temples~\cite{KnapitschPZK17} dataset (bounded ver.).}
    The results are averaged over 5 scenes.}
    \label{tab:bd_tt_summary}
    \end{subtable}
    \vspace{-.5em}
    \caption{
    {\bf Results on bounded inward scenes.}
    We only compare with the recent fast convergence approaches.
    Our small and large models use $160^3$ and $256^3$ voxels respectively, and both are measured on an RTX 2080Ti GPU.
    Results breakdown and rendered videos: \url{https://sunset1995.github.io/dvgo/results.html}.
    }
    \label{tab:bd_inward}
    % \vspace{-1em}
\end{table}

%%%%%%%%%%%%%%%%%%%%%%%%%

\subsection{Forward-facing scenes}
\paragraph{Points parameterization and sampling.}
We use NeRF's parameterization to warp the unbounded forward-facing frustum to a bounded volume.
In this case, the dense voxel grid allocation is similar to the multiplane images (MPIs)~\cite{ZhouTFFS18}, where we place $D$ RGB-density images at fixed depths, each with $X\times Z$ resolution.
Every ray is traced from the first to the $D$-th images with a step size of $s$ image, \ie, there are $2D-1$ sampled points if $s=0.5$.

\paragraph{Implementation details.}
The number of depth layers is $D{=}256$ each with $XZ{=}384^2$ number of voxels.
Sampling step size is $s{=}1.0$ layer.
The TV loss weights are $10^{-5}$ for the density grid and $10^{-6}$ for the feature grid; the distortion loss weight is $10^{-2}$.

\paragraph{Results.}
We compare DVGOv2 with the recent fast convergence approaches in \cref{tab:llff_summary}.
DVGOv2 shows comparable quality using less training time on the lowest spec GPU.
We also see that the efficient distortion loss makes our training faster and achieves better quality, thanks to the compactness encouraged by the loss.
We also note that we achieve similar performance using a much lower grid resolution.
This is perhaps due to the other challenges in the LLFF dataset (\eg, fewer training views with some multi-view inconsistency due to real-world capturing), which hinders the gain by using higher grid resolution.

\begin{table}[h!]
    \centering
    \begin{tabular}{@{}l@{\hskip 3pt}|@{\hskip 3pt}c@{\hskip 8pt}c@{\hskip 8pt}c@{\hskip 8pt}c@{}}
    \hline
    Method & Tr. time & PSNR{\scriptsize$\uparrow$} & SSIM{\scriptsize$\uparrow$} & LPIPS{\scriptsize(VGG)$\downarrow$} \\
    \hline\hline
    {\small Plenoxels~\cite{YuFTCR22}} & \gray{24.2m} & 26.29 & \gold{0.839} & 0.210 \\
    {\small TensoRF (S)~\cite{ChenXGYS22}} & \gray{19.7m} & \silver{26.51} & 0.832 & 0.217 \\
    {\small TensoRF (L)~\cite{ChenXGYS22}} & \gray{25.7m} & \gold{26.73} & \gold{0.839} & \silver{0.204} \\
    % DVGOv2 (S) & 7.0 & 25.85 & 0.825 & 0.218 \\
    % DVGOv2 (L) & 10.7 & 26.12 & \brown{0.833} & \silver{0.207} \\
    {\small DVGOv2 {\footnotesize w/o $\mathcal{L}_{\text{dist}}$}} & 13.9m & 26.24 & 0.833 & \silver{0.204} \\
    {\small DVGOv2} & 10.9m & \brown{26.34} & \brown{0.838} & \gold{0.197} \\
    \hline
    \end{tabular}
    \vspace{-1em}
    \caption[]{
        {\bf Results on LLFF~\cite{MildenhallSOKRNK19} dataset.\footnotemark[3]}
        The results are averaged over 8 scenes.
        The effective grid resolution is {\footnotesize $1408\times 1156\times 128$} for Plenoxels and {\footnotesize $640^3$} for TensoRF, while ours is about {\footnotesize $384^2\times 256$}.
    }
    \label{tab:llff_summary}
    \vspace{-1em}
\end{table}

%%%%%%%%%%%%%%%%%%%%%%%%%

\subsection{Unbounded inward-facing scenes}
\paragraph{Points parameterization and sampling.}
We adapt mip-NeRF 360~\cite{BarronMVSH22} parameterization for the unbounded 360 scenes, which is
\begin{equation} \label{eq:ub}
    % \vspace{-.5em}
    \bf{x}' = \begin{cases}
        \bf{x} , & \|{\bf x}\|_p \leq 1~; \\
        \left(1+b-\frac{b}{\|{\bf x}\|_p}\right) \left(\frac{\bf{x}}{\|{\bf x}\|_p}\right) , & \|{\bf x}\|_p > 1 ~. \\ 
    \end{cases}
    % \vspace{-.5em}
\end{equation}
We first rotate the world coordinate to align the first two PCA directions of camera positions with grid's XY axis, which slightly improves our results.
The world coordinate is then shifted to align with cameras' centroid and scaled to cover all cameras near planes in a unit sphere.
We allocate a cuboid voxel grid centered at the origin with length $2{+}2b$.
The hyperparameter $b{>}0$ controls the proportion of voxel grid points allocated to the background.
The sampling step size is measured directly on the contracted space.
The original $p{=}2$ wastes around 50\% of the grid points (a sphere in a cube), so we also try $p{=}\infty$ to make a cuboid contracted space.

\paragraph{Implementation details.}
The grid resolution is $320^3$.
We set $\alpha^{\text{(init)}}=10^{-4}$~\cite{SunSC22} with $0.5$ voxel step size.
The TV loss weights are $10^{-6}$ for the density grid and $10^{-7}$ for the feature grid; the distortion loss weight is $10^{-2}$.

\paragraph{Results.}
We show our results on the unbounded inward-facing scenes in \cref{tab:tt_ub_summary} and \cref{tab:mipnerf360_summary}.
On the Tanks\&Temples~\cite{KnapitschPZK17} dataset (\cref{tab:tt_ub_summary}), we achieve comparable SSIM and LPIPS to NeRF++, while our quality is behind the Plenoxels due to the grid resolution limited by the dense grid.
On the newly released mip-NeRF 360 dataset (\cref{tab:mipnerf360_summary}), we achieve NeRF comparable PSNR and SSIM, while our LPIPS is still far behind.
One improvement is scaling the grid resolution as we only use $320^3$ voxels, while advanced data structures~\cite{YuFTCR22,mueller2022instant,ChenXGYS22} are necessary in this case.
However, DVGOv2 could still be a good starting point for its simplicity (Pytorch, dense grid) with reasonable quality.
Using a cuboid contracted space significantly improves our results on mip-nerf 360 dataset, while it degrades the results on Tanks\&Temples dataset perhaps due to the photometric inconsistency problem in the dataset.
Again, the distortion loss~\cite{BarronMVSH22} with our efficient realization improves our quality and training speed.

\begin{table}[h!]
    \centering
    \begin{tabular}{@{}l@{\hskip 3pt}|@{\hskip 3pt}c@{\hskip 8pt}c@{\hskip 8pt}c@{\hskip 8pt}c@{}}
    \hline
    Method & Tr. time & PSNR{\scriptsize$\uparrow$} & SSIM{\scriptsize$\uparrow$} & LPIPS{\scriptsize(VGG)$\downarrow$} \\
    \hline\hline
    {\small NeRF++~\cite{ZhangRSK20}} & \gray{hours} & \gold{20.49} & 0.648 & \brown{0.478} \\
    {\small Plenoxels~\cite{YuFTCR22}} & \gray{27.3m} & \silver{20.40} & \gold{0.696} & \gold{0.420} \\
    % DVGOv2 (S) & 18.0 & 19.98 & 0.637 & 0.509 \\
    {\small DVGOv2 {\footnotesize w/o $\mathcal{L}_{\text{dist}}$}} & 22.1m & 20.08 & \brown{0.649} & 0.495 \\
    {\small DVGOv2} & 16.0m & \brown{20.10} & \silver{0.653} & \silver{0.477} \\
    \hline
    \end{tabular}
    \vspace{-1em}
    \caption[]{{\bf Results on Tanks\&Temples~\cite{KnapitschPZK17} dataset.\footnotemark[3]}
    The results are averaged over the 4 scenes organoized by NeRF++~\cite{ZhangRSK20}.
    The grid resolution of Plenoxels is $640^3$ for foreground and $2048\times1024\times64$ for background, while ours is a single {\footnotesize $320^3$} grid shared by foreground and background.
    }
    \label{tab:tt_ub_summary}
    \vspace{-1em}
\end{table}

\begin{table}[h!]
    \centering
    \begin{tabular}{@{}l@{\hskip 3pt}|@{\hskip 3pt}c@{\hskip 8pt}c@{\hskip 8pt}c@{\hskip 8pt}c@{}}
    \hline
    Method & Tr. time & PSNR{\scriptsize$\uparrow$} & SSIM{\scriptsize$\uparrow$} & LPIPS{\scriptsize(VGG)$\downarrow$} \\
    \hline\hline
    {\small NeRF~\cite{MildenhallSTBRN20}} & \gray{hours} & 24.85 & 0.659 & \brown{0.426} \\
    {\small NeRF++~\cite{ZhangRSK20}} & \gray{hours} & \silver{26.21} & \silver{0.729} & \silver{0.348} \\
    {\small mip-NeRF 360~\cite{BarronMVSH22}} & \gray{hours} & \gold{28.94} & \gold{0.837} & \gold{0.208} \\
    {\small DVGOv2 {\footnotesize w/o $\mathcal{L}_{\text{dist}}$}} & 16.4m & 24.73 & 0.663 & 0.465 \\
    {\small DVGOv2 $p{=}2$} & 13.2m & 24.80 & 0.659 & 0.468 \\
    {\small DVGOv2 $p{=}\infty$} & 14.0m & 25.24 & 0.680 & 0.446 \\
    {\small DVGOv2 $p{=}\infty ^{(*)}$} & 15.6m & \brown{25.42} & \brown{0.695} & 0.429 \\
    \hline
    \multicolumn{5}{l}{{\footnotesize $^{(*)}$ Longer grid scaling and decaying schedule.}}
    \end{tabular}
    \vspace{-1em}
    \caption[]{{\bf Results on mip-NeRF-360~\cite{BarronMVSH22} dataset.\footnotemark[3]}
    The results are averaged over {\bf only the publicly available 7 scenes}.}
    \label{tab:mipnerf360_summary}
    \vspace{-1em}
\end{table}

\footnotetext[3]{\scriptsize Results breakdown and rendered videos: \url{sunset1995.github.io/dvgo/results.html}}

\clearpage

%%%%%%%%% REFERENCES
{\small
\bibliographystyle{ieee_fullname}
\bibliography{neural_rendering}
}
% %%%%%%%%%
% \clearpage\newpage

\end{document}